\documentstyle[osa,graphicx]{revtex}

\begin{document}
	
\title{Suppression of Classical and Quantum Radiation Pressure Noise via
Electro-Optic Feedback}

\author{Ben C. Buchler, Malcolm B. Gray, Daniel A. Shaddock, Timothy C. Ralph and
David E. McClelland}

\address{Department of Physics, Faculty of Science\\
The Australian National University\\
Australian Capital Territory 0200, Australia.}
\maketitle
\begin{center}
	To be published in Optics Letters
\end{center}
\begin{abstract}
We present theoretical results that demonstrate a new technique to be 
used to improve the sensitivity of thermal noise measurements: 
intra-cavity intensity stabilisation.  It is demonstrated that 
electro-optic feedback can be used to reduce intra-cavity intensity 
fluctuations, and the consequent radiation pressure fluctuations, by a 
factor of two below the quantum noise limit.  We show that this is 
achievable in the presence of large classical intensity fluctuations 
on the incident laser beam. The benefits of this scheme are a consequence of 
the sub-Poissonian intensity statistics of the field inside a feedback 
loop, and the quantum non-demolition nature of radiation pressure noise
as a readout system for the intra-cavity intensity fluctuations.
\end{abstract}

Interferometric gravity wave detectors measure the relative motion of 
suspended mirrors. Any fluctuation in mirror position, other than that 
due to a gravity wave, is undesirable. One significant source of 
mirror motion is that due to thermal noise \cite{saulson}. This is the 
result of resonant modes of the the mechanical mirror and suspension 
system being excited by their intrinsic thermal energy. Methods of 
measuring thermal noise are important for the development of low noise 
mirror and suspension systems.

\begin{figure}[h!]
\centering          
\includegraphics[width=10cm]{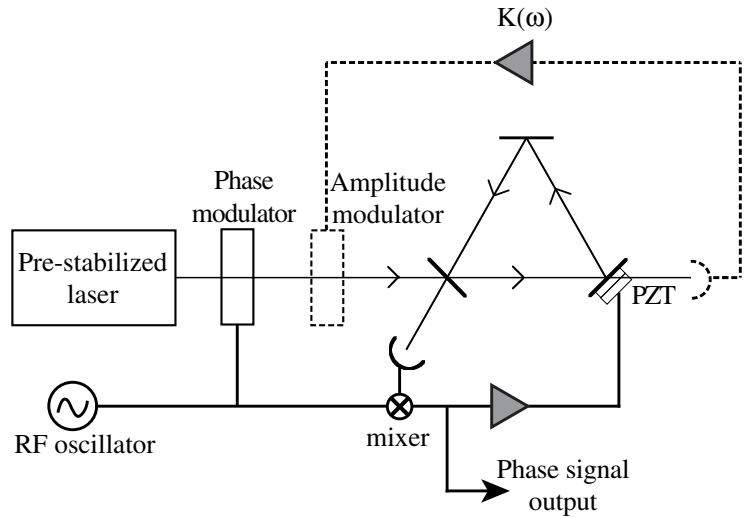}   
\caption{The experimental arrangement for a thermal noise measurement. 
The dashed part of the figure indicates the feedback loop}
\label{experiment} 
\end{figure}

We will consider the thermal noise measurement system shown in figure 
\ref{experiment}. For clarity, the figure shows a ring cavity used for 
thermal noise measurement while in practice a linear Fabry-Perot 
cavity will be used. The cavity is locked to the laser using 
reflection locking \cite{drever}. The laser is frequency stabilized 
using a reference cavity which is long compared to the test cavity. 
Above the bandwidth of this locking system, (experimentally this could 
be above 10Hz) the error signal will represent a measurement of the 
phase noise of the reflected beam. With the laser sufficiently 
stabilized by the reference cavity, the phase noise measured by the 
error signal will be due to cavity detuning. In the absence of seismic 
and acoustic noise, the detuning is due to thermal and radiation 
pressure noise. It is the radiation pressure noise which will limit 
the performance of this thermal noise measurement system.

An impedance-matched cavity is considered for two reasons. Firstly, it 
gives optimum phase sensitivity to the reflection locking system. 
Secondly, with no reflected carrier intensity the locking error 
signal represents a quantum limited measure of the phase noise of the 
reflected field.   A fully reflective cavity, on the other 
hand, would require a local oscillator beam to make a quantum limited 
measurement of the phase noise of the reflected beam. Phase noise 
measurement on the reflected beam is therefore considerably simplified 
for an impedance matched cavity.

To find the limiting performance of this system, we will model it 
using the input/output formulation of optical cavities \cite{Qoptics}. 
After linearization, this method provides frequency domain expressions 
for the amplitude and phase quadrature fluctuation operators which in 
turn yield expressions for the phase and intensity noise spectra 
\cite{ral96}. Assuming the experiment is isolated from all seismic and 
acoustic sources, the dimensionless time domain cavity mode 
annihilation operator $\hat{a}$ will be given by
\begin{eqnarray}
\dot{\hat{a}}&=&-\left(\kappa + \imath \delta
\hat{X}_{{\Delta}_{T}} + \imath \delta \hat{X}_{\Delta_{R}}\right)\hat{a}
\nonumber \\
&& + \sqrt{2 \kappa_{in}} \hat{A}_{in} + \sqrt{2 \kappa_{out}}
\delta \hat{A}_{\nu} + \sqrt{2 \kappa_{L}}
\delta \hat{A}_{L},
\label{timeeq}
\end{eqnarray}
where $\delta \hat{X}_{{\Delta}_{T}}$ is the quadrature operator for 
the cavity detuning due to the thermal noise of the cavity and $\delta 
\hat{X}_{\Delta_{R}}$ is the quadrature operator for detuning due to 
radiation pressure. These operators have units of $1/sec$. The term 
$\hat{A}_{in}$ is the field operator for the laser input to the 
cavity; $\delta \hat{A}_{\nu}$ is the field operator for the vacuum 
entering at the cavity output mirror and $\delta \hat{A}_{L}$ is the 
field operator for the vacuum noise due to intra-cavity loss. These 
terms have units of $1/\sqrt{sec}$. The $\delta$ indicates that an 
operator represents a small fluctuation and the expectation value of 
the operator is 0. The rate constants in Eq. \ref{timeeq} are: 
$\kappa_{in}$, the cavity loss through the front cavity mirror; 
$\kappa_{out}$, the cavity loss through the rear cavity mirror and 
$\kappa_{L}$, the intra-cavity loss. The sum of these three rates is 
the total cavity loss rate, $\kappa$. We are interested in the 
fluctuations of the cavity system about steady state, so the operators 
in the above equation which are not already fluctuation operators are 
written in the form $\hat{a}=\alpha+\delta \hat{a}$ and 
$\hat{A}_{in}=A_{in}+\delta\hat{A}_{in}$ where $\alpha$ and $A_{in}$ 
are the (assumed real) steady state values for the cavity mode 
amplitude and the input field amplitude respectively. These are 
substituted into Eq. \ref{timeeq} which is then linearized by 
retaining only first order fluctuation terms. Taking the Fourier 
transform allows solution for the frequency domain cavity mode 
fluctuation operator, $\delta \tilde{a}$, where tildes will be used to 
indicate Fourier domain operators.

Using the relation $\delta \tilde{A}_{ref}=\sqrt{2 \kappa_{in}}\delta 
\tilde{a} -\delta \tilde{A}_{in}$ \cite{Coll} between the reflected 
field fluctuations $\delta \tilde{A}_{ref}$ and the cavity 
mode, we may find an expression for the phase quadrature operator of the 
reflected field fluctuations $\delta \tilde{X}_{A_{ref}}^{-}$ where 
the phase quadrature of an operator $\tilde{b}$ is defined as $\delta 
\tilde{X}_{b}^{-}=i(\delta \tilde{b} -\delta \tilde{b}^{\dagger})$. 
The RF phase noise spectrum, $V_{ref}^{-}(\omega)$, is evaluated using 
the relation $\delta(\omega-\omega')V_{ref}^{-}(\omega)=\left<\delta 
\tilde{X}_{A_{ref}}^{-}(\omega) \delta 
\tilde{X}_{A_{ref}}^{-^{*}}(\omega')\right>$ giving
\begin{equation}
V_{ref}^{-}(\omega)=1+\frac{8\kappa_{in}\alpha^{2}\left(V_{\Delta_{R}}(\omega)
+V_{\Delta_{T}}(\omega)\right)+\left((2\kappa_{in}-\kappa)^{2}+\omega^{2}\right)
(V_{in}^{-}(\omega)-1)}{\omega^{2}+\kappa^{2}}
 \label{phase}
\end{equation}
We have used the fact that the variance of the quantum vacuum fields 
is normalized to 1.  The phase noise of the reflected field therefore 
represents a measurement of the thermal noise spectrum of the cavity, 
$V_{\Delta_{T}}(\omega)$, provided the other noise sources are small 
enough compared to the thermal noise present in the system.  The phase 
noise of the input field, $V_{in}^{-}(\omega)$, can be brought close 
to $1$ (i.e.  a coherent state) by using a suitable reference cavity 
for laser stabilisation.  This noise source can therefore be 
neglected.  The only other noise source we can control is the 
radiation pressure noise $V_{\Delta_{R}}(\omega)$.  We will assume 
that there is a transfer function, $F(\omega)$, between the 
intra-cavity intensity fluctuations, $V_{a}(\omega)$, and the phase 
noise due to the radiation pressure fluctuations \cite{saulson} so 
that
\begin{equation}V_{\Delta_{R}}(\omega)=F(\omega)V_{a}(\omega).
\label{radp}\end{equation}

Logically, an effective method for improving the measurement of 
thermal noise is to reduce the intra-cavity intensity noise. Using a 
derivation analogous to that given above for the phase noise, the 
frequency domain amplitude quadrature noise operator (defined as 
$\delta \tilde{a}+ \delta \tilde{a}^{\dagger}$) of the cavity mode is 
given by
\begin{equation}
\delta \tilde{X}_{a}(\omega)=\frac{\sqrt{2\kappa_{in}}\delta
\tilde{X}_{A_{in}}+\sqrt{2\kappa_{out}} \delta
\tilde{X}_{A_{\nu}}+\sqrt{2\kappa_{L}} \delta
\tilde{X}_{A_{L}}}{\kappa+\imath \omega}.
\label{Xa}\end{equation}
By evaluating $\delta(\omega-\omega')V_{a}(\omega)=\left<\delta 
\tilde{X}_{a}(\omega) \delta \tilde{X}_{a}^{*}(\omega')\right>$ the 
cavity mode intensity noise is found to be
\begin{equation}
V_{a}(\omega)=\frac{2\kappa_{in}V_{in}+2\kappa_{out}+2\kappa_{L}}{\kappa^{2}+
\omega^{2}}.
\label{Va}\end{equation}
Assuming we are well within the cavity line-width ($\omega<<\kappa$),
the best we can do without using a squeezed state of light is a
coherent state, i.e. $V_{in}(\omega)=1$,  in which case we find
\begin{equation}
V_{a}=\frac{2}{\kappa}. \label{coh}
\end{equation}

The effect of a feedback system, shown by the dotted parts of figure
\ref{experiment} will now be considered.  The light transmitted by the
cavity is detected and fed-back to an amplitude modulator in the input
field.  We model its effect by modifying the form of the input field
amplitude quadrature fluctuation \cite{tau,feedback} so that
\begin{equation}
\delta \tilde{X}_{A_{in}} \rightarrow \delta
\tilde{X}_{A_{in}}-\delta \tilde{X}_{R}, \label{feedsub}
\end{equation}
where $\delta \tilde{X}_{R}$ are the extra amplitude
fluctuations added by the modulator. This form of the feedback is
justified provided the amplitude modulator is just that, and adds no
additional phase signal. The minus sign in Eq. \ref{feedsub} is chosen so that
later results agree with conventional control theory notation.  The
operator $\delta \tilde{X}_{R}$ will be a function of the field
detected by the feedback detector, which in turn is a function of the
intra-cavity amplitude fluctuations.  The operator $\delta
\tilde{X}_{R}$ may be written as
\begin{equation}
\delta \tilde{X}_{R}=\sqrt{2\kappa_{out} \eta} K(\omega)
\delta \tilde{X}_{a}- K(\omega)\left(\sqrt{\eta}\delta
\tilde{X}_{A_{\nu}}+\sqrt{1-\eta}\delta \tilde{X}_{A_{D}}\right),
\label{fform}
\end{equation}
where $\delta \tilde{X}_{A_{D}}$ is the vacuum noise operator
introduced by an imperfect detector with efficiency $\eta$ and
$K(\omega)$ is the transfer function of the electronics in the
feedback loop. The phase of $K(\omega)$ must be engineered to ensure 
feedback loop stability. Substituting Equations \ref{feedsub} and \ref{fform}
into Equation \ref{Xa} and re-evaluating the intra-cavity intensity
noise spectrum we find that feedback gives
\begin{equation}
V^{f}_{a}=\frac{2 \kappa_{in}V_{in} + 2 \kappa
_{in}(1-\eta)|K(\omega)|^{2} +
\left|\sqrt{2 \eta \kappa_{in}} K(\omega)+\sqrt{2
\kappa_{out}}\right|^{2}
+2\kappa_{L}}{\left|\kappa +\imath\omega +2
\sqrt{\kappa_{out}\kappa_{in}\eta} K(\omega)\right|^{2}}
\label{invar}
\end{equation}
In the limit of low frequency ($\omega<<\kappa$) and high gain 
($|K(\omega)|\rightarrow \infty$) we find that
\begin{equation}
V^{f}_{a}=\frac{1}{2 \eta \kappa_{out}} \label{Vfeed}
\end{equation}
The contributions to this spectrum in the limit of high gain come only 
from the vacuum noise entering the back of the cavity, $\delta 
\tilde{X}_{A_{\nu}}$, and vacuum noise due to inefficient 
photo-detection, $\delta \tilde{X}_{A_{D}}$. Noise due to 
intra-cavity loss and the input laser intensity 
noise are fully suppressed by the feedback loop. Comparing the 
intra-cavity intensity noise with feedback (Eq. \ref{Vfeed}) to that 
found for a coherent input in Eq. \ref{coh} we find
\begin{equation}
\frac{V^{f}_{a}}{V_{a}}=\frac{\kappa}{4\eta \kappa_{out}} \label{ans}
\end{equation}
If we consider an impedance matched cavity, with
$\kappa_{in}=\kappa_{out}=\kappa/2$ and $\kappa_{L}=0$ we obtain
\begin{equation}
\frac{V^{f}_{a}}{V_{a}}=\frac{1}{2\eta}. \label{imp}
\end{equation}
Equation \ref{imp} shows that in the case of an impedance matched 
cavity and 100\% efficient photodetection, the system provides a 
factor of two reduction of intra-cavity intensity noise compared to 
that found for a coherent input. A factor of two reduction in 
intra-cavity intensity noise naturally leads, via Eq. \ref{radp}, to a 
corresponding reduction in radiation pressure noise therefore 
enhancing the performance of our thermal noise measurement. This 
improvement is independent of any classical intensity noise present on 
the input laser beam because, in the high gain limit, all classical 
noise is suppressed by the feedback loop.

To highlight the applicability of this technique to real experiments 
we present a sample calculation based on measured parameters of a 
realistic Fabry-Perot cavity. With a transmission efficiency of 95\%, 
and 1\% power reflection, we can infer the cavity parameters to be 
$\kappa_{in}=\kappa_{out}=4.9\kappa_{L}$. Given detectors with a 
quantum efficiency of 91\%, equation \ref{ans} gives an intra-cavity 
intensity noise (and therefore radiation pressure noise) suppression 
of 2.2dB relative to a coherent laser input. For the diode pumped 
Nd:YAG lasers commonly used in the authors' laboratory, intensity 
fluctuations at low frequency are typically 60dB above the standard 
quantum limit \cite{lightwave}. In order to satisfy the high gain 
limit, a loop gain of approximately 80dB or greater is required. This 
is readily achieved using high speed analogue electronics as part of a 
well designed feedback loop.

The above description shows the power fluctuations of the intra-cavity 
field to be ``squashed'' below those found for a coherent state. 
However, the intra-cavity field is not squeezed. An intensity squeezed 
state has increased phase noise whereas feedback has no such phase 
noise penalty. Other authors have described in detail the peculiar 
nature of fields inside a feedback loop. Shapiro \cite{shap} showed 
explicitly that the in-loop field does not obey free field 
commutators. Although it has also been shown by Haus \cite{hau86} that 
the commutator relations for a cavity mode inside a feedback loop are 
preserved. To see the intensity noise squashing of light inside a feedback loop 
is not easy. In fact it requires the use of a quantum non-demolition 
(\textit{QND}) measurement of the in-loop field \cite{wis94}. Our 
scheme benefits from the in-loop squashing because radiation 
pressure noise is a \emph{QND} probe of the intra-cavity intensity 
noise \cite{Jac94}. The suppressed in-loop noise is therefore evident 
on the readout of the radiation pressure noise present on the phase 
quadrature of the field reflected from the cavity.

Finally, it is worth noting that within the bandwidth of the intensity 
feedback loop, the out-of-loop field (in this case this is the 
reflected field from the cavity) will always have super-Poissonian 
intensity noise due to the quantum measurement penalty 
\cite{feedback,hau86,wis94,wis942}. This \emph{intensity} noise 
penalty does not, however, affect the readout of the \emph{phase} 
noise by the reflection locking technique. This would not be the case 
if this experiment were to be performed with real amplitude squeezed 
light.  One would obtain a reduction in the radiation pressure noise, 
however the increased phase noise which comes with amplitude squeezed light 
would impinge upon the phase measurement of the reflected field.

\acknowledgments
The authors would like to thank H.-A Bachor for useful discussions.  
This work was supported by the Australian Research Council.

\end{document}